\documentclass[twocolumn,showpacs,preprintnumbers,amsmath,amssymb, floatfix, prb]{revtex4} 
\usepackage{graphicx}
\usepackage{subfigure}  
\usepackage{umlaut}  
\begin{document}  
\title{Ehrlich-Schwoebel Effect for Organic Molecules: Direct Calculation of the Step Edge Barrier using Empirical Potentials}  
\author{M. Fendrich}  
\affiliation{  
Fachbereich Physik, Universität Duisburg-Essen,\\ D-47048 Duisburg, Germany,\\ email: markus.fendrich@uni-due.de\\}  
\author{J. Krug}   
\affiliation{  
Institut für Theoretische Physik, Universität zu Köln,\\ D-50937 Köln, Germany,\\ e-mail: krug@thp.uni-koeln.de\\}  
\begin{abstract}  
The step edge barrier of a prototypical organic semiconductor molecule, 3,4,9,10-perylene-tetracaboxylic-dianhydride (PTCDA) has been analysed by means of calculations based on emperical potentials. The minimum energy path (MEP) has been calculated for a single molecule on a substrate of three molecular layers between equivalent minimum energy positions within two neighboring unit cells. To determine the step edge 
barrier, we have calculated the MEP over a step to a fourth layer of molecules. We found energy barriers of $E_D= 80$ meV for in-layer diffusion and $E_S = 750$ meV for step  
crossing, indicating a strong Ehrlich-Schwoebel effect for PTCDA.  
\end{abstract}  
\maketitle  
  
In the last decades, there has been a growing interest in electronic devices built with organic semiconductor molecules. Organic molecular devices, in contrast to present-day electronics based on silicon, offer a higher mechanical flexibility and lower production costs. However, this field of research still faces many open questions and future challenges. One of them is the understanding of the growth of organic crystals,\cite{Schreiber2004} as the quality of an organic crystal has a strong influence on device properties such as, e.g., the charge carrier mobility.  
 
An important factor in inorganic crystal growth kinetics is the Ehrlich-Schwoebel (ES) effect,\cite{Schwoebel66JAP37_3682,Ehrlich66JCP44_1039} which refers to the additional energy barrier experienced by 
an atom or molecule crossing a step edge. The strength of the step edge barrier controls the amount of mass transport between different crystalline layers, and hence determines the temperature dependence 
of the \textit{kinetic} growth mode (two-dimensional layer-by-layer or three-dimensional mound growth) in situations where energetic driving forces shaping the surface morphology can be neglected. 
\cite{Michely2004}   
 
There is considerable indirect evidence that points to the presence of a step edge barrier for   
prototypical organic semiconductor molecules like 3,4,9,10-perylene-tetracaboxylic-dianhydride (PTCDA)\cite{wagner04OE5_35,Kilian04SS573_359,Krause2004,Kunstmann05PRB71_121403(R), 
Mativetsky07Nanotechnology18_105303} and pentacene, 
\cite{mzh01nature412_517,Ruiz2004,zorba06PRB74_245410} and the ES effect has been invoked in several simulation studies of 
organic film growth \cite{Krause2004,zorba06PRB74_245410,Stephan2000}. Very recently a quantitative estimate for the step edge barrier on pentacene has been obtained 
from an analysis of experimental layer coverages \cite{Mayer06PRB73_205307}.  
So far, however, the existence of step edge barriers for organic molecules has not been established 
on the molecular level.   
The problem in theoretical studies is the high computational effort: Due to the size of the relevant systems ($\approx$50 atoms per molecule for relevant molecules, $\approx$100 molecules in a model molecular crystal), \textit{ab initio} methods are not suitable. Empirical potentials offer an opportunity to avoid these problems, if one accepts the corresponding loss in accuracy.  
The capability of empirical potentials for growth studies of organic thin films has been demonstrated in recent publications.\cite{Mannsfeld05PRL94_056104,Mannsfeld04PRB69_075461,Verlaak2007}  
\par
\begin{figure}[t]  
 \includegraphics[width=7.5cm]{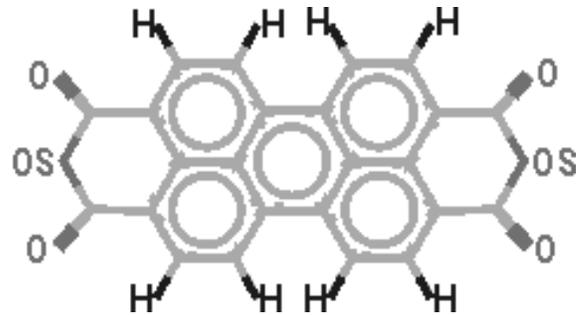}  
\caption{PTCDA molecule (C$_{24}$O$_6$H$_8$). The size of the molecule (with van der Waals radii) is 14.2 \AA $\times$ 9.2 \AA. The two oxygen atoms denoted 
by OS have slightly different parameters in the AMBER force field than those denoted by O.}  
\label{ptcda}  
\end{figure}

 
In this paper, we present calculations of the diffusion and step edge barriers for a PTCDA molecule (see Fig. \ref{ptcda}) on a PTCDA crystal. Energies have been calculated using empirical potentials; the general method presented here should also apply to other molecular systems.    
  
To model a PTCDA single crystal surface, a rigid cell of three layers of PTCDA molecules in $\alpha$-configuration was set up, using the crystallographic data by Ogawa et al.\cite{ogawa99ACB55_123}. The optimum postion for a single rigid molecule on the topmost layer was determined using a gradient search algorithm. The diffusion barrier $E_D$ was then calculated by finding the minimum energy path (MEP) between minimum positions within two neighboring PTCDA unit cells, using the nudged elastic band algorithm.\cite{Henkelman00JCP113_9978}  
Potential energies have been calculated by means of empirical pair potentials of Lennard-Jones (LJ) type. Parameters from the AMBER molecular force field \cite{Weiner84JACS106_765} were used. AMBER uses a 6-12 LJ potential with the energy given by  
\[  
E_{LJ}\;=\;\sum_{i,j}\left[\frac{A_{ij}}{R_{ij}^{12}}-\frac{B_{ij}}{R_{ij}^6}\right],  
\]     
where $i$ and $j$  
denote all pairs of atoms with distance $R_{ij}$ in the PTCDA substrate ($i$) and the single PTCDA molecule on top ($j$).  
$A_{ij}$ and $B_{ij}$ are specified by the parameters $r^*$ and $\epsilon$ which are given for each atom type (see table \ref{rstareps}). The combination rules are  
\begin{eqnarray}  
A_{ij}\;=\;\left(\frac{r_i^*}{2}+\frac{r_j^*}{2}\right)^{12}\;\sqrt{\epsilon_i\epsilon_j}\nonumber\\  
B_{ij}\;=\;2\left(\frac{r_i^*}{2}+\frac{r_j^*}{2}\right)^6\;\sqrt{\epsilon_i\epsilon_j}\nonumber  
\end{eqnarray}   
PTCDA is a polar molecule, therefore we had to account for electrostatic interactions. The electrostatic energy has been calculated by assigning partial charges to all atoms by a Mulliken method, using the commercial software Hyperchem. A dielectric constant of $\epsilon_r$ = 3.2 was used, which was found as an average value for bulk PTCDA.\cite{Forrest97CR_1793} The total electrostatic energy is then calculated by summing up all pairs of charged atoms. This method has recently been applied successfully to find the energy barriers for another organic molecular system.\cite{Fendrich06PRB73_115433}  
  
\begin{table}[h]  
\caption{Parameters for the atoms in the PTCDA system, as given in the AMBER force field}  
\label{rstareps}  
\begin{ruledtabular}  
\begin{tabular}{cccc}  
  atom & AMBER atom type & $r^*$[nm] & $\epsilon$[kcal/mol]\\  
  \hline  
  C & C/CA & 0.185 & 0.12\\  
  O & O & 0.160 & 0.20\\  
  O & OS & 0.160 & 0.15\\  
  H & H & 0.100 & 0.02\\  
\end{tabular}  
\end{ruledtabular}  
\end{table}  
  
\begin{figure}[t]  
 \includegraphics[width=8.7cm]{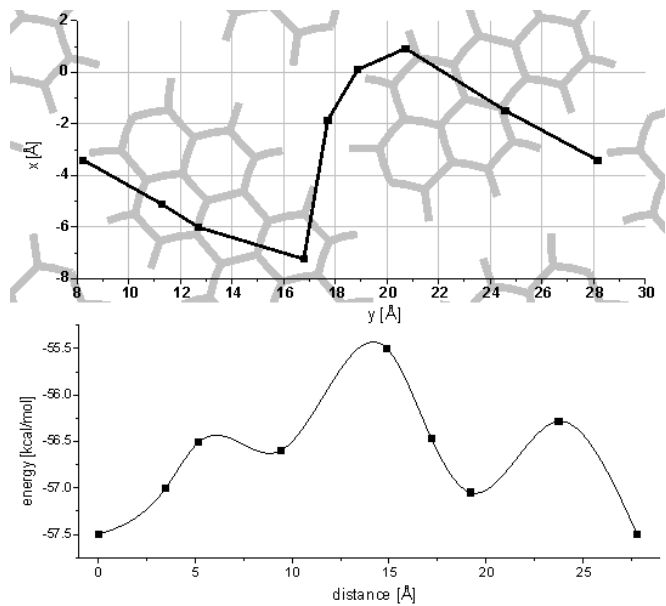}  
\caption{Minimum energy path (MEP) for the center of mass (top) and potential energy along this path (bottom) for a PTCDA molecule on top of three layers of PTCDA in the (102) plane of the bulk $\alpha$ configuration. The molecules of the topmost layer are shown in grayshade. The MEP reveals a diffusion barrier for a single molecule of $E_D$ = 80 meV.}  
\label{3layers}  
\end{figure}

\begin{figure}[h]  
 \includegraphics[width= 8.7cm]{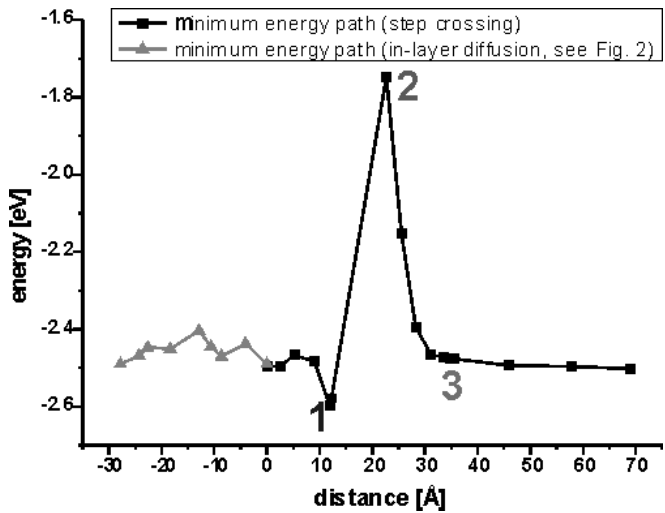}  
\caption{(Color online) Potential energy along the MEP for a PTCDA molecule moving over a PTCDA step edge. The step edge barrier is $E_S$ = 750 meV. The MEP for in-layer diffusion is also drawn. Three important positions for a PTCDA molecule: (1) built into the step, (2) on the edge of the molecular step, (3) on the upper terrace.}  
\label{step}  
\end{figure}

The calculation of the MEP for a PTCDA molecule from its minimum position within a PTCDA unit cell to the corresponding position in the neighboring unit cell reveals a diffusion barrier of $E_D$ = 80 meV. The path and the potential energy along this path are shown in Fig. \ref{3layers}. Note that the minimum energy position of the molecule is not the same as the position in the bulk unit cell; this can be explained as the energy landscape for a single molecule should differ from that of a molecular layer.  
In a second calculation we determined the step edge barrier for a molecule. The model system was extended by a fourth layer area on top of the three layers. Then, again, the MEP for a PTCDA molecule over this single molecular step was calculated, revealing a fairly high barrier. Figure \ref{step} shows the potential energy along the MEP over the step edge. The energy drops shortly before the step: The molecule gains energy when it is built in at the step. To go over the step, the molecule then has to overcome an energetic barrier of $E_S$ = 750 meV.  
  
To put these energy barriers into perspective, it is useful to introduce the concept of an \textit{onset} temperature\cite{Michely2004} for a given kinetic 
process, which can be defined (somewhat arbitrarily) as the temperature at which the process takes place at a rate of 1 $s^{-1}$. 
Assuming a standard attempt frequency \cite{Michely2004,Fendrich06PRB73_115433} of $10^{12} \, s^{-1}$, our estimated energy barriers imply an onset temperature  
of 33 K for diffusion and 315 K for step edge crossing. This indicates that the PTCDA molecules should be highly mobile at any temperature of interest, 
whereas interlayer transport is completely frozen: Even at 500 K, the ratio of the interlayer mobility to the intralayer diffusion rate is only $e^{-(E_S - E_D)/k_{\mathrm{B}} T} \approx  
2 \times 10^{-7}$, which implies that a molecule has to interrogate the step edge $5 \times 10^6$ times before descending.  
 
These predictions are consistent with growth experiments: When evaporated on an insulating KBr surface,\cite{Kunstmann05PRB71_121403(R),Mativetsky07Nanotechnology18_105303} PTCDA exhibits  
three-dimensional growth. Even at low coverages, crystalline islands with a minimum height of 3 layers grow, and no wetting layers of PTCDA can be seen.  
On metal surfaces, the behavior is different.\cite{wagner04OE5_35,Kilian04SS573_359} 
In this case  Stranski-Krastanov-type behavior is found, in that the growth of three dimensional islands sets in after the formation of two complete layers of molecules.  
The epitaxial structure of the first two layers, however, indicates a strong interaction with the substrate, which is not included in the calculations.  
The initial layer-by-layer growth does therefore not contradict our theoretical study. 
  
The small diffusion barrier we found also suggests the presence of mobile PTCDA molecules if experiments are performed at room temperature. Experimental evidence for this was found in scanning probe microscopy (SPM) experiments: When high resolution images of the topmost layer are taken, the images often become streaky. This is, e.g., described in ref. \cite{Mativetsky07Nanotechnology18_105303}; the authors explain this behavior with mobile molecules that are diffusing on the topmost layer.   
 
\par Due to the high step edge barrier, molecules should be trapped on the topmost layer at room temperature. On the other hand, the small diffusion barrier might prevent molecules on top from nucleating and forming an incomplete additional layer. Indeed, in most SPM experiments, the top layers of PTCDA crystals seem complete at room temperature, as the mobile molecules cannot be resolved. In a recent experiment, \cite{Fendrich_unpub} PTCDA molecules were found to form small clusters of molecules on top of crystallites when the sample is cooled down to 80 K, while at room temperature, the topmost layers seemed flat and complete.  

To summarize, 
we presented a study based on empiricial potentials to determine the diffusion barrier $E_D$ and the step edge barrier $E_S$ for an important organic molecule, PTCDA. On top of a cell of molecules in the bulk configuration, the MEPs for a single molecule between minimum energy positions within neighboring unit cells and over a molecular step edge were calculated, yielding the estimates $E_D$ = 80 meV and $E_S$ = 750 meV. Although a very coarse model was set up, these results are consistent with experimental data. The accuracy of our estimates of $E_D$ and $E_S$ is difficult to assess. However, we note that the binding energy of a PTCDA molecule found using this approach, $E_B^{th} \approx 2.5$ eV, is in close agreement with thermal desorption spectroscopy experiments\cite{Wagner_unpub}, which yield 
a binding energy of $E_B^{exp}$ = 2.2 eV.  
The method described in this paper should therefore apply to many molecular systems and help to predict and explain effects in organic molecular crystal growth.  
  
Financial support was granted by the Deutsche Foschungsgemeinschaft through SFB616 ''Energy Dissipation at Surfaces''.  
We are grateful to T. Michely and T. Wagner for fruitful discussions.

\end{document}